\documentclass[aps,pra,twocolumn,groupedaddress,showpacs]{revtex4}

\usepackage{hyperref, soul}
\usepackage{graphicx}
\usepackage{amsfonts,amsmath,amssymb}
\usepackage{appendix}
\usepackage[latin1]{inputenc}

\graphicspath{fig/}

\begin{document}

\title{Quantum correlations by four--wave mixing in an atomic vapor in a non-amplifying regime: a quantum beam splitter for photons.}

\author{Quentin Glorieux}

\email{quentin.glorieux@univ-paris-diderot.fr}

\author{ Luca Guidoni}

\author{Samuel Guibal}

\author{Jean-Pierre Likforman}

\author{Thomas Coudreau}

\affiliation{Univ Paris Diderot, Sorbonne Paris Cité, Laboratoire Matériaux et
Phénomènes Quantiques, UMR 7162 CNRS, F-75205 Paris, France}


\begin{abstract} 
We study the generation of intensity quantum correlations using four--wave mixing in a rubidium vapor. The absence of cavity in these experiments allows to deal with several spatial modes simultaneously.
In the standard, amplifying, configuration, we measure relative intensity squeezing up to 9.2~dB below the standard quantum limit.
We also theoretically identify and experimentally demonstrate an original regime where, despite no overall amplification, quantum correlations are generated.
In this regime a four--wave mixing set--up can therefore play the role of a photonic beam splitter with non--classical properties, \textit{i.e.} a device that splits a coherent state input into two quantum correlated beams.
\end{abstract}

\maketitle

Non--classical ''intense'' beam have been widely studied in a large variety of contexts including potential applications to quantum information protocols \cite{Braunstein05,Sangouard10}, fundamental issues in quantum mechanics such as entanglement and non--locality \cite{Reid09}, quantum imaging  \cite{QuantumImaging} or enhancement of the sensitivity of gravitational wave interferometers \cite{Caves81}. 
Quantum correlated beams are usually obtained through optical non--linear effects described as $\chi^{(2)}$ or  $\chi^{(3)}$ non linearities present in a variety of media (see \cite{BachorBook} for a review).
In this paper, we study the generation of quantum correlation by using four--wave mixing (4WM) in a hot atomic vapor.

Based on $\chi^{(3)}$ non linearity, 4WM is well known to generate intense non classical beams~\cite{Yuen79,Slusher1985,Shelby86,Maeda87}. 
However, over the last 20 years, attention has been focused mainly on $\chi^{(2)}$ media \cite{Wu87,Heidmann87,Laurat05,Mehmet10} mainly because of their low losses (availability of high quality optical crystals).
On the contrary, in hot vapors the presence of an atomic resonance enhances the non-linearity but also usually increases the losses.
Yet, recently, it was shown that non--degenerate 4WM in atomic vapors can produce very large amounts of quantum correlations between intense beams \cite{McCormick2007,McCormick08,GlorieuxSPIE10}.
Such a set--up has a significant advantage over $\chi^{(2)}$ media in that it does not require an optical cavity to enhance the nonlinearity and the related quantum effects. 
This is particularly important in the case of quantum imaging where spatially multimode quantum effects are involved \cite{QuantumImaging,Boyer08}.
Furthermore, the generated beams directly match the atomic resonance frequency of an atom--based quantum memory, a key requirement for quantum communications \cite{Sangouard10}.

As noted above, the large nonlinear and quantum effects observed in 4WM originate from the presence of an atomic resonance. 
This resonance also induces incoherent effects, most notably absorption and spontaneous emission, which, in general, decrease the degree of quantum correlations. 
These possible drawbacks are often reduced by increasing the detuning from resonance, resulting in an overall amplification of the probe and conjugate beams.
However, as we show, there exists a regime where quantum correlations can be observed despite the fact that the probe beam is de-amplified by propagation through the atomic vapor.
In this regime, a 4WM setup then behaves as a beam--splitter separating an incoming beam in two different beams, without overall amplification.

However, when the input beam is in a coherent state, the two output states are quantum correlated : we thus call this new device a \textit{quantum beam--splitter for photons}.
On the two input ports of the device are respectively sent the vacuum state and a coherent state and through the two output ports are emitted the quantum correlated states.
This denomination omits the role of the pump which is crucial in this scheme as naturally no classical beam splitter can generate non classical states starting from coherent states.
The simplest way to model theoretically such a device is to chain an ideal linear phase--insensitive amplifier with a partially transmitting medium.
Despite the introduction of large losses, up to a level that cancels the gain, we show that quantum correlated beams can be generated in such a configuration.
We then introduce the gemellity \cite{Treps05}, a criterion well adapted to describe experiments with non balanced beams.
\begin{figure}[h]
  \centerline{\includegraphics[width=1\columnwidth]{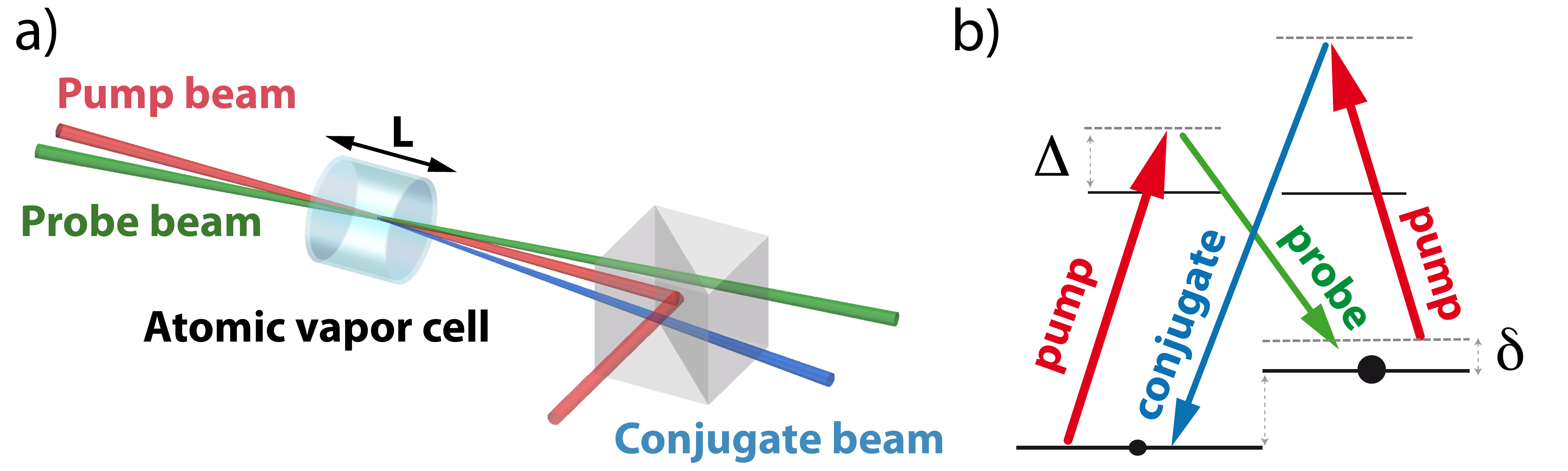}}
  \caption{(Color Online) a) 4WM in hot atomic vapor schematic setup. 
  b) Relevant levels of the Rb D1 line described as a double-$\Lambda$ system. $\Delta$ is the so called one--photon detuning and $\delta$ the two--photon detuning.}
  \label{fig:levels}
\end{figure}

We demonstrate, using a microscopic model\cite{GlorieuxPRA10}, that 4WM in a hot atomic vapor can efficiently implement a  \textit{quantum beam--splitter} and we show that the limit for the maximum gemellity predicted in the linear amplifier model can be theoretically exceeded in this new regime.
Finally we test these predictions experimentally.

\section{4WM in the amplifying regime}
\label{sec:ampl-regime}

The experiment is based on \cite{McCormick2007} and is described in detail in \cite{GlorieuxSPIE10} so that we only recall here its main features.
A linearly polarized intense pump beam, frequency locked near the $^{85}$Rb $D_1$ line, is mixed with an orthogonally polarized weak probe beam inside an isotopically pure  cell of length $L$.
The relevant levels are shown in Fig.~\ref{fig:levels}, a.
At the output of the cell, due to 4WM, the probe beam is amplified and a conjugate beam is generated (see Fig.~\ref{fig:levels}, b).
After filtering out the pump beam with a polarizing beam splitter, intensity correlations between the probe and conjugate beams are measured by a pair of high quantum-efficiency photodiodes coupled to a spectrum analyzer.

A high gain can be observed for a relatively large set of experimental parameters.
The use of a heated cell yields a large number of atoms: for a temperature $T$ ranging from 100$^\circ$C to 150$^\circ$C, the atomic density $\mathcal{N}$ calculated from the Clausius--Clapeyron formula \cite{Alcock84} varies from 6$\times 10^{12}~$cm$^{-3}$ to $10^{14}~$cm$^{-3}$. 
Thus, the equivalent optical depth, $\mathcal N \sigma L$ varies between $5\times 10^3$ and $10^5$  where $\sigma$ is the atomic cross section for the 5$S^{1/2}$ $\to$ 5$P^{1/2}$ transition in $^{85}$Rb. 
These atoms interact with beams close to resonance : the single photon detuning $\Delta$ is typically 1~GHz (on the order of the Doppler broadening) while the two--photon detuning $\delta$ is smaller than 10~MHz. 

\begin{figure}[h]
  \centering
  \includegraphics[width=.7\columnwidth]{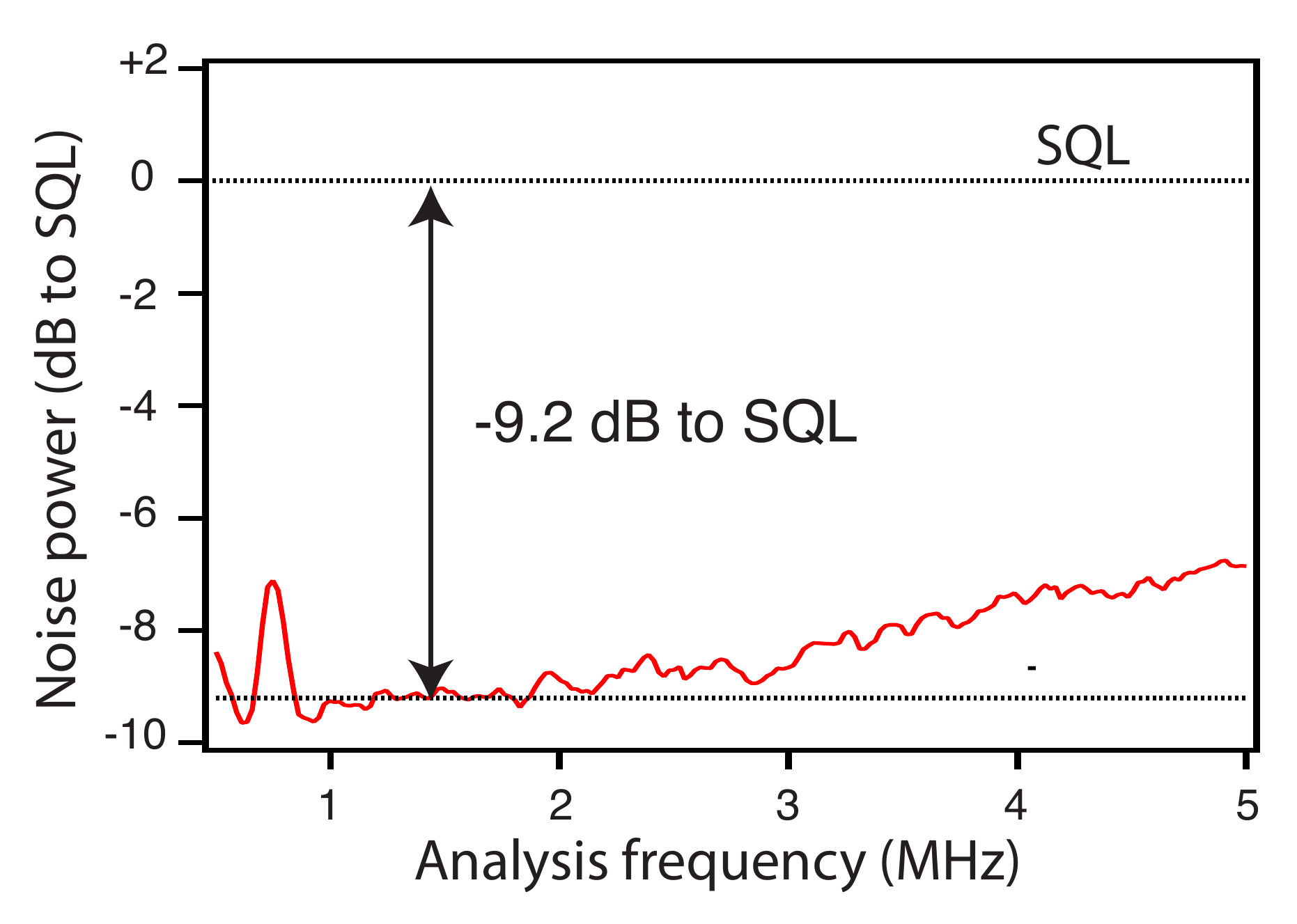}
  \caption{(Color Online) Noise power of the intensity difference between the probe and
    conjugate beams as a function of the frequency after correcting for the
    electronic noise. A reduction of 9.2~dB$\pm$0.5~dB below the SQL is reached at 1~ MHz analysis frequency. }
  \label{fig:record}
\end{figure}

Within these domains of parameters, explored systematically \cite{GlorieuxSPIE10}, we have identified an optimal noise reduction regime.
For $\Delta=+750$~ MHz, $\delta=+6$~ MHz, $T = 118^\circ $C, $P_{pump}=1200$~mW (corresponding to a Rabi frequency $\Omega= 1$~GHz), gain on the incoming probe beam up to 20 can be observed. 
In these conditions, Fig.~\ref{fig:record} shows the noise power of the intensity difference of the probe and conjugate as a function of the analysis frequency after correcting for the electronic noise:significant noise reduction is observed in the range 500~kHz to 5~MHz and with a maximal noise reduction of 9.2~dB$\pm$0.5~dB below the SQL is observed between 1 and 2~MHz. This value is slightly larger than the best results obtained to date  with 4WM \cite{McCormick08} and very close to those obtained with OPOs \cite{Laurat05}. The matching of the atomic resonance of Rb turns this setup into an ideal source of non-classical light to interact with Rb vapor quantum memory \cite{PL1,PL2}.

\section{Quantum beam splitter regime}
\label{sec:de-amplif}

In the previously described regime,the larger was the gain, the larger were the quantum correlations.
However, we will show here that this not a necessary condition and that one can, somewhat counter-intuitively, observe significant quantum correlations in the absence of overall gain.

\subsection{Ideal linear amplifier model}
\label{sec:ideal-ampl}

In an ideal phase--insensitive amplifier, an input probe beam is amplified while a conjugate beam is generated.
At the output of the amplifier, neglecting the contribution of the noise to the average number of photons, the probe beam has an intensity $G I_0$ and the conjugate beam has an intensity $(G-1)I_0$ where $G$ denotes the gain and $I_0$ the input probe beam intensity.
Taking into account the ideal character of the amplifier, no noise is added and the intensity difference at the output has a noise ratio $1/(2G-1)$ with respect to the input.
For probe and conjugate at the input respectively coherent and vacuum states, this noise ratio is equal to the quantum correlations at the output of the amplifier.
If we now extend this model by including losses at the output of the medium on the probe and/or conjugate beams, one would expect a reduction in these correlations as it is well known that losses are detrimental to squeezing.
Let us recall that this is not always the case as the beams intensity are not balanced: a small amount of extra losses on the probe beam will tend to make the two beams more balanced and thus improve the noise reduction on the intensity difference as noted \textit{e.g.} in \cite{Jasperse11}.

{Contrary to the case of Optical Parametric Oscillators above threshold \cite{Laurat05}, 4WM naturally generates} unbalanced beams.
{Unbalanced beams may exhibit strong quantum correlations but the measurement of the noise on the intensity difference is not an ideal criterion in this case. 
It is useful to introduce the gemellity} $\mathcal{G}$ \cite{Laurat03,Laurat04,Treps05,Laurat05b} defined by :
\begin{equation}
\mathcal{G}=\frac{F_a+F_b}{2}-\sqrt{C_{ab}^2F_aF_b+\left(\frac{F_a-Fb}{2}\right)^2},
\end{equation}
where $F_{i}=\langle \hat X_{i}\hat X_{i}\rangle$ with $i$ used for $a$ (probe) and $b$ (conjugate), $C_{ab}= \frac{\langle \hat X_{a}\hat X_{b}\rangle}{\sqrt{F_aF_b}}$ and $\hat X_{i}$ is the amplitude quadrature of the related field as defined in \cite{Treps05}. 
{In case of balanced beams the gemellity is equal to the normalized noise on the difference between the fluctuations of the two measurements~: $\mathcal{G}=\frac{\langle (\hat X_{i}-\hat X_{j})^2\rangle}{2} $, which is the quantitative measure of the maximal ``non--classicality'' that can be extracted from the correlated beams \cite{Treps05}}.
For balanced beams, such as the ones produced in the limit of infinite gain, this value is equal to the standard criterion, namely the intensity noise difference.
In the conditions above where the intensity difference noise is -9.2~dB, the noise on the individual beams are $F_a = F_b$ = +12~dB at 1~MHz yielding a gemellity $\mathcal{G}$ = -9.8~dB $\pm$ 0.5~dB. This value is comparable with record values measured with an OPO above threshold \cite{Laurat05}and moreover a large number of spatial modes (estimated to 100 in this particular configuration) are squeezed simultaneously \cite{Boyer08} .

Using this criteria and introducing losses on the probe ($T_a$) and conjugate ($T_b$) beams so that the overall transmission is equal to one ($T_a G + T_b (G-1)=1$), it is straightforward to show that there always exists a region in the parameter space where a gemellity lower that one is expected.
To our knowledge, this phenomenon, albeit simple, has neither been discussed nor observed.
The larger quantum correlations reachable with no overall amplification corresponds to the situation of a gain $G=1.23$, a transmission of 0.62 on the probe beam and perfect transmission on the conjugate beam. This configuration gives the	limit for the gemellity reachable by this simple model : $\mathcal{G} = -2.8$~dB.

\subsection{Microscopic model}
\label{sec:micro-model}

To investigate further this effect, we have studied the 4WM process using a microscopic model based on the cold--atom model described extensively in \cite{GlorieuxPRA10}.
This model assumes the simplified double--$\Lambda$ level structure  of Fig.~\ref{fig:levels}, right.
The Heisenberg--Langevin approach is used to obtain the relevant classical quantities (probe gain $G_a$, conjugate gain $G_b$ defined with respect to $I_0$) as well as the quantum properties of the output beams. 
In particular, it is possible to calculate noise spectra that allow for quantifying quantum correlations both in terms of intensity--difference noise $S_{N_-}$ and for the unbalanced case in terms of gemellity  $\mathcal{G}$.
In the regime of high amplification previously described, this model is in good quantitative agreement with the measured correlations \cite{GlorieuxPhD}.
Exploring the parameters space in this model, we have found a new region where the 4WM process generates quantum correlations in the absence of overall amplification. 
This regime is therefore very similar to the linear amplifier model followed by a lossy medium described above.
Nevertheless, the microscopic model predicts that in this regime, the gemellity can be significantly enhanced in contrast to the linear model and exceeds the -2.8 dB limit discussed previously.

\begin{figure}
\centering
\includegraphics[width=0.79\columnwidth]{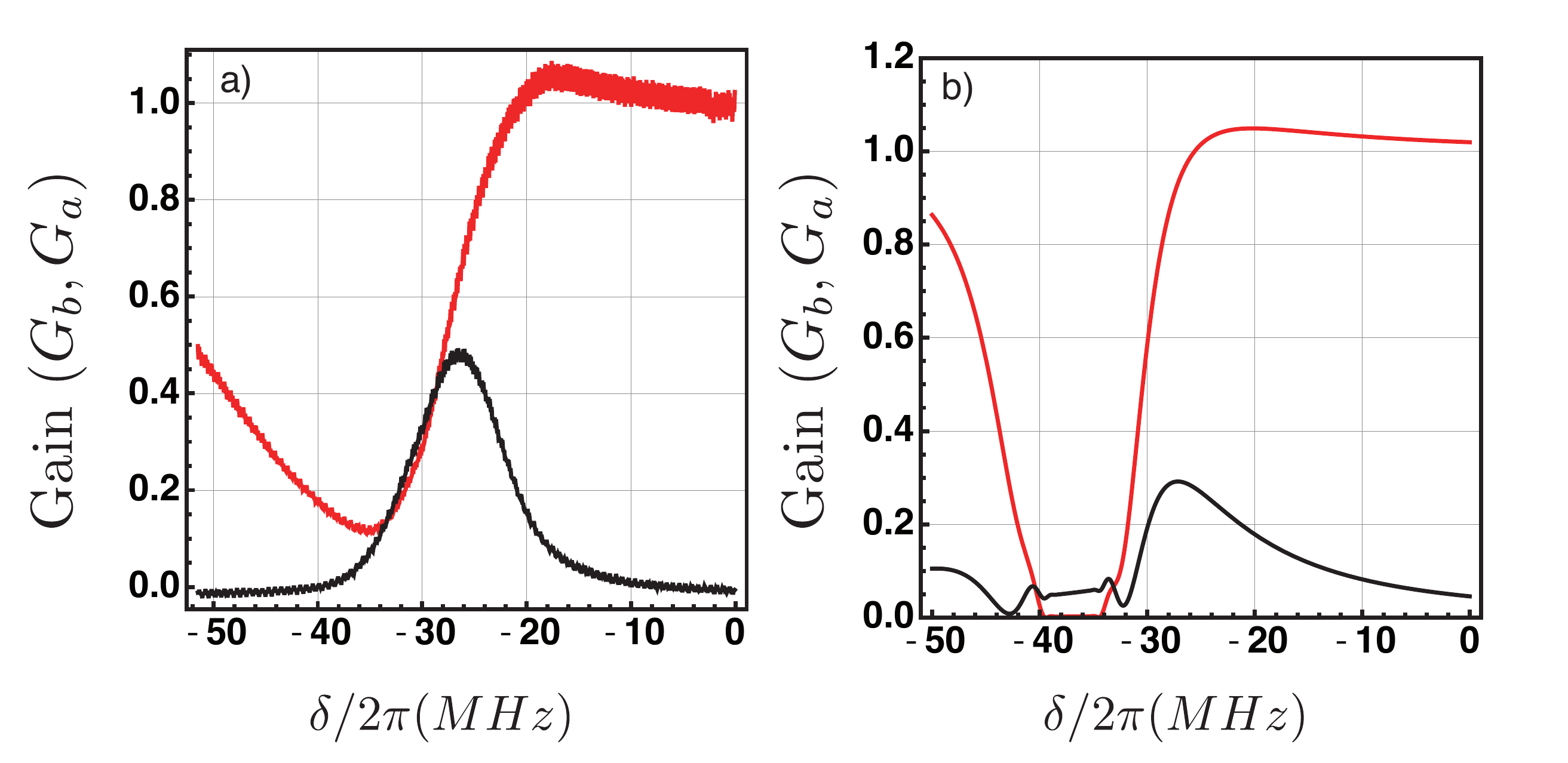} 
\caption{(Color Online) Theoretically predicted (left) and experimentally measured (right) gain for the probe beam ($G_a$, ) and conjugate beam ($G_b$, black) as a function of the two--photon detuning $\delta$. The parameters  used in the simulations are : , optical depth $\mathcal N \alpha L=500$, Pump Rabi frequency $\Omega$ = 0.42~GHz, single photon detuning $\Delta/2\pi$=0.8~GHz. 
Measured parameters are : pump power $P$=0.6~W ($\Omega /2 \pi=0.4$~GHz), T=95$^\circ$C, single photon detuning $\Delta/2\pi$=0.8~GHz.\label{fig:qbs-gain-exp}}
\end{figure}

Let us start by presenting the classical behavior of the probe and conjugate beams in the region of interest of parameter space (theoretical data are compared to the experimental results).
We plot in Fig.~\ref{fig:qbs-gain-exp}, the gain for the two fields as a function the two--photon detuning $\delta$.
The main difference with respect to the high gain parameter region is the choice of the atomic density (experimentally driven by the temperature).
The large gain results of Fig.~\ref{fig:record} were obtained for a temperature of 118~$^\circ$C while the curves in Fig.~\ref{fig:qbs-gain-exp} are obtained for  $T=95^\circ$C.
The approximately one order of magnitude lower optical density, together with the different choice of $\delta$ and $\Delta$, explain the drastic reduction of $G_a$ and $G_b$.
"Beam--splitter" regime is obtained near the two-photon resonance, where $G_a$ goes to zero due to a Raman process involving a probe and a pump photon\cite{GlorieuxPRA10}.
Due to the pump--induced AC-Stark shift, this two--photon resonance is shifted to negative values of $\delta$ and its exact position depends on the one-photon detuning $\Delta$ and on the pump Rabi frequency $\Omega$.
Within a very narrow region of parameter space, the sum of the two beams output intensities becomes slightly smaller or almost equal to the input probe intensity.
{It is interesting to note that for potential applications this very narrow feature could be considered as a limitation.
Notwithstanding we have verified numerically, by changing simultaneously $\Delta$, $\Omega$ and the optical depth $\mathcal N \alpha L$, that the detuning for which this system exhibits the behavior of a quantum beam splitter can be tuned over more than 100 MHz.}
As already remarked in \cite{GlorieuxPRA10}, we note that, despite the fact that the model is based on a cold atom sample, it yields without any adjustable parameter a qualitative agreement with the experimental data obtained in a hot vapor.

\begin{figure}
\centering
\includegraphics[width=0.75\columnwidth]{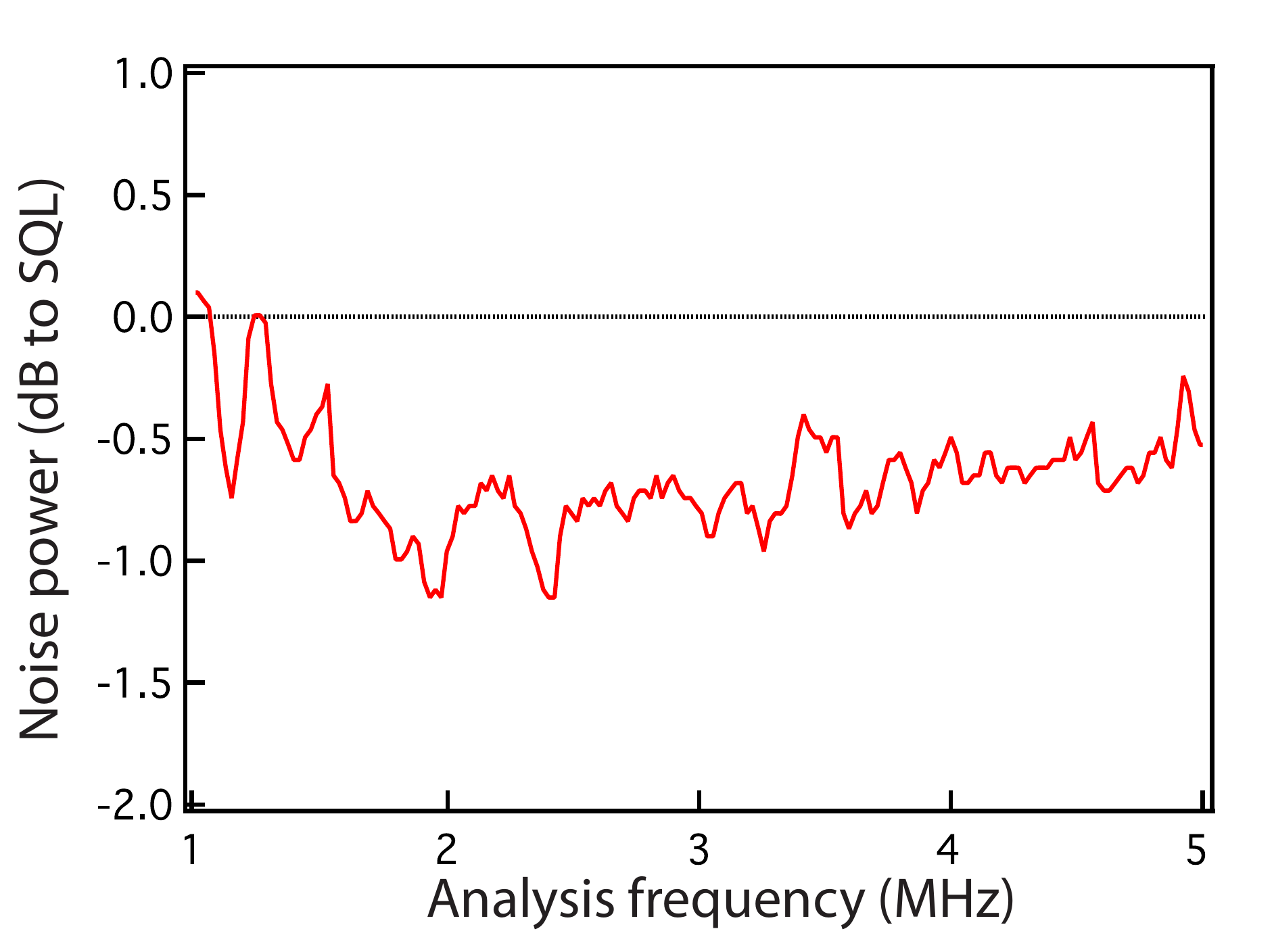}
\caption{(Color Online) Quantum intensity correlations between the probe and conjugate beams as a function of the analysis frequency with the same parameters as above and $\delta/2\pi$=-52~MHz. \label{fig:qbs-quantum}}
\end{figure}

\subsection{Demonstration of the quantum beam splitter}

Motivated by these theoretical predictions, we have experimentally investigated this original regime.
{We plot in Fig.~\ref{fig:qbs-quantum} the experimentally measured intensity difference noise as a function of the analysis frequency $\omega$.}
We observe significant quantum correlations, down to 1.0$\pm 0.2$~dB below the SQL around an analysis frequency of 1 MHz.
At the same time, the power of the two beams normalized to the probe input power is measured to be 0.65 and 0.35 for the probe and conjugate respectively.  
This demonstrates clearly the behavior of a quantum beam splitter for photons where one laser beam is split into two beams without gain but generating quantum correlations.
We note that the measured noise reduction is slightly smaller than the one predicted theoretically (Fig.~\ref{fig:qbs-quantum}): this discrepancy can be attributed to the fact that the model is based on a cold atomic sample, far from the experimental regime.

In this situation, $\mathcal{G}$ can be calculated to compare it to the theoretical limit of the linear amplifier model.
By measuring the noise on the two individuals beams respectively equal to + 3 and + 2 dB for probe and conjugate, we obtain a value of the gemellity equal to $\mathcal{G}=-1.8\pm 0.5$~dB.
This value does not exceed the maximum limit of -2.8 dB predicted by the linear amplifier model.
As previously noted, the theoretical model does not take into account the velocity distribution of the atoms, thus time transit effects and Doppler broadening which are expected to play a detrimental role. 
This can explain why the linear amplifier model limit cannot be reached in this configuration whereas the microscopic model predicts that gemellities better than $\mathcal{G}$=-3.2~dB can be obtained with the above parameters and an optical depth, $\mathcal{N} \sigma L =1500$.

We have thus shown firstly that generating quantum correlations does not require overall amplification and, secondly, that the ideal linear amplifier is not the ideal device to perform this operation but that 4WM in atomic vapours presents an interesting avenue in this context.
This setup could also be used as the input beam splitter introducing quantum correlations for an original version of the Mach-Zender interferometer as well as in a so-called SU(1,1) interferometer \cite{Yurke86}.

\section{Conclusion}
\label{sec:conclusion}

We have studied the production of quantum correlated beams in four--wave mixing in a $^{85}$Rb cell.
{First, we have identified and experimentally realized  an optimal regime in the high gain region where intensity--difference noise down to -9.2~dB below the standard quantum limit (gemellity $\mathcal{G}=-9.8~dB$) have been measured.}
This result is important in the domain of quantum communications where both large non--classical effects and the availability of an atom--based storage media form strong requirements \cite{Sangouard10,PL1,PL2}.

We have also predicted and observed an original regime where quantum correlations are present despite significant losses on the probe beam.
This regime is of particular interest, because it can occur in a situation in which the sum of the two output beam intensities is smaller or equal to the input probe intensity.
Therefore the atomic medium controlled by the pump laser acts like a beam splitter device that creates quantum correlations (quantum beam splitter).
Although this effect could in principle be observed with an ideal amplifier, it is to our knowledge the first demonstration of it.
In this context, we have discussed the use of the gemellity criterion more appropriate in the case of unbalanced beams produced by four--wave mixing.
Finally, a microscopic model allowed us to demonstrate that 4WM in the quantum beam splitter regime can beat theoretically the limit of quantum correlations predicted by the model of a linear amplifier followed by a lossy medium.

In our experiment, with a hot atomic vapor, a value of  $\mathcal{G}=-1.8\pm 0.5$~dB has been reported. 
Although the parameter values required to beat the linear amplifier model limit are presently beyond reach of experiments performed with cold atoms, our model provides an interesting avenue to surpass this limit using hot or cold atoms.

\section*{Acknowledgements}
We thank E. Arimondo, R. Dubessy and P. Lett for fruitful discussions.
We thank N. Treps for the loan of high--efficiency photodiodes.

\end{document}